\documentclass[12pt,preprint]{aastex}

\newcommand{\psr}{PSR~J1909$-$3744}

\shorttitle{PSR~J1909$-$3744, a Binary Millisecond Pulsar}
\shortauthors{Jacoby et al.}

\begin{document}

\title{PSR~J1909$-$3744, a Binary Millisecond Pulsar with a Very Small Duty Cycle}

\author{B. A. Jacoby\altaffilmark{1}, M. Bailes\altaffilmark{2},
M. H. van Kerkwijk\altaffilmark{3}, S. Ord\altaffilmark{2},
A. Hotan\altaffilmark{2, 4}, S. R. Kulkarni\altaffilmark{1}, and
S. B. Anderson\altaffilmark{1}}

\altaffiltext{1}{Department of Astronomy, California Institute of
Technology, MS 105-24, Pasadena, CA 91125; baj@astro.caltech.edu,
srk@astro.caltech.edu, sba@astro.caltech.edu.}

\altaffiltext{2}{Centre for Astrophysics and Supercomputing, Swinburne
University of Technology, P.O. Box 218, Hawthorn, VIC 31122,
Australia; mbailes@swin.edu.au, sord@swin.edu.au, ahotan@swin.edu.au.}

\altaffiltext{3}{Department of Astronomy and Astrophysics, University
of Toronto, 60 St. George Street, Toronto, ON M5S 3H8, Canada;
mhvk@astro.utoronto.ca.}

\altaffiltext{4}{Australia Telescope National Facility, CSIRO, P.O. Box 76, Epping, NSW 1710, Australia.}

\begin{abstract}
We report the discovery of \psr, a 2.95~millisecond pulsar in a nearly
circular 1.53~day orbit.  Its narrow pulse width of 43\,$\mu$s allows
pulse arrival times to be determined with great accuracy.  We have
spectroscopically identified the companion as a moderately hot ($T
\approx 8500$\,K) white dwarf with strong absorption lines.  Radial
velocity measurements of the companion will yield the mass ratio of
the system.  Our timing data suggest the presence of Shapiro delay; we
expect that further timing observations, combined with the mass ratio,
will allow the first accurate determination of a millisecond pulsar
mass.  We have measured the timing parallax and proper motion for this
pulsar which indicate a transverse velocity of
$140^{+80}_{-40}$\,km\,s$^{-1}$.  This pulsar's stunningly narrow
pulse profile makes it an excellent candidate for precision timing
experiments that attempt to detect low frequency gravitational waves
from coalescing supermassive black hole binaries.
\end{abstract}


\section{INTRODUCTION}

Binary radio pulsars provide a rich laboratory for a wide range of
physical inquiry, including neutron star masses and the evolution of
binary systems.  The pulsars with the best-determined masses all have
eccentric relativistic orbits, spin periods of tens of milliseconds,
and masses clustered very tightly around 1.35\,M$_{\odot}$
\citep{tc99, bok+03}.  It is generally thought that millisecond
pulsars ($P \le 10$\,ms) should have greater mass than these
longer-period pulsars due to a larger accreted mass.  There is indeed
a statistical suggestion that this is the case, but to date no precise
millisecond pulsar mass measurements exist.  Currently, the
best-determined millisecond pulsar masses are $m_{\rm p} = 1.57^{+
0.12}_{- 0.11}\,{\rm M}_\odot$ for PSR~B1855+09 \citep{nss03} and
$m_{\rm p} = 1.58 \pm 0.18\,{\rm M}_\odot$ for PSR~J0437$-$4715
\citep{vbb+01}.

\psr\ was discovered during the Swinburne High Latitude Pulsar Survey,
a recently-completed pulsar search using the 13-beam multibeam
receiver on the 64-m Parkes radio telescope \nocite{jac03} (Jacoby
2003; Jacoby et al. in preparation).  This survey is similar to, and an
extension of, the highly-successful Swinburne Intermediate Latitude
Pulsar Survey \citep{ebvb01, eb01, eb01b}.  The pulsar was discovered
in three adjacent survey beams observed on 2001 January 25 and
confirmed on 2001 May 26.  Due to the pulsar's strong broad-band
scintillation, the initial position inferred from the discovery data
was in error by nearly a full beamwidth; it was not until we obtained
a phase-connected timing solution that we realized the correct
position.

\begin{figure}
\plotone{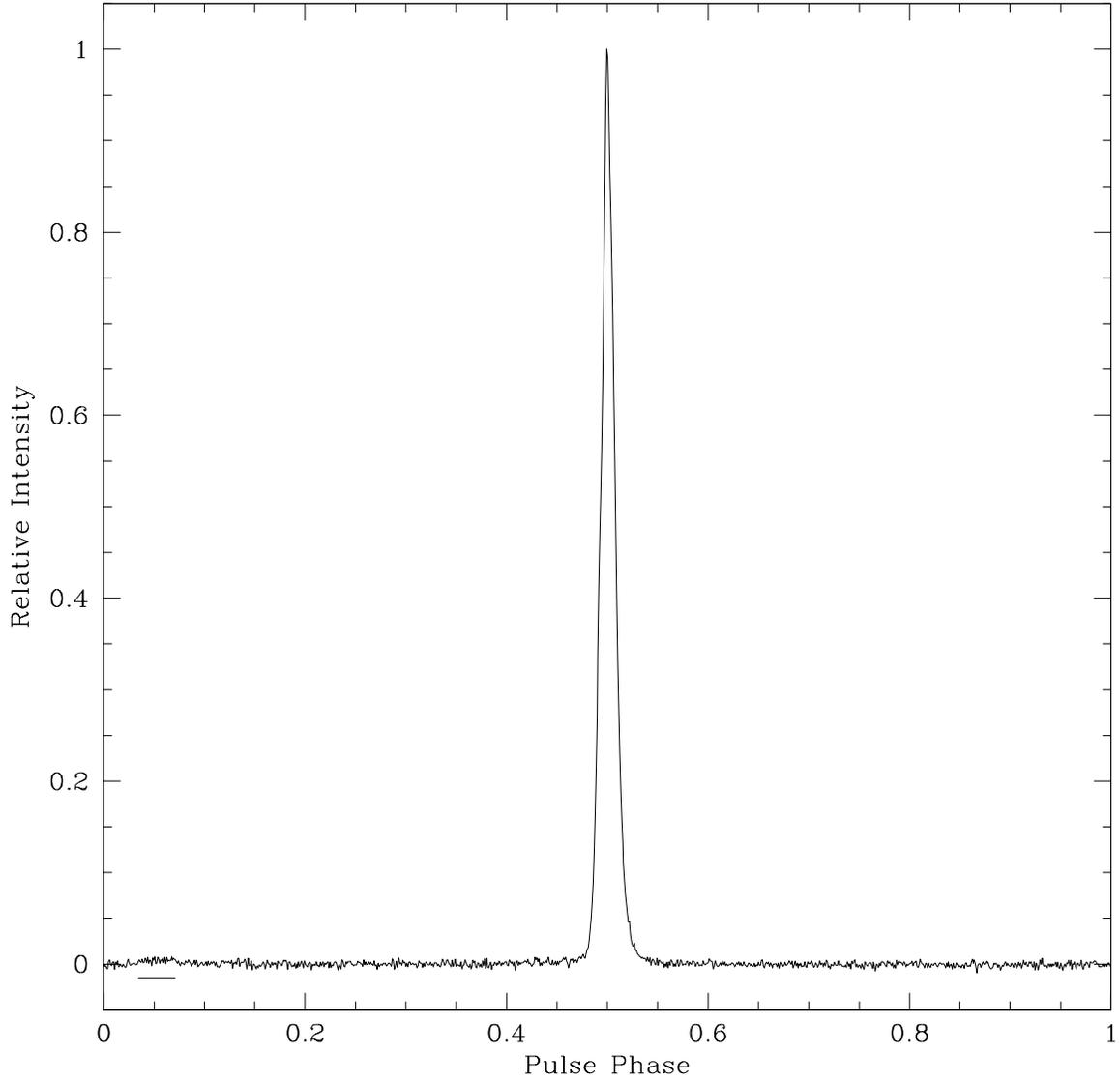}
\caption{Coherently-dedispersed average pulse profile of \psr\ at
1341~MHz as measured by the CPSR2 baseband system. Instrumental
smearing is just 2$\mu$s. The location of a very weak interpulse is
indicated by the underscore at left.}
\label{fig:1909_profile}
\end{figure}

Among the properties that can make a pulsar useful for precision
timing experiments are a bright, narrow pulse profile, extremely
stable rotation, proximity to the earth, and a companion star which
can be detected and studied.  \psr\ appears to possess all of these
desirable traits.  Nearly all of the pulsar's flux comes in a single
narrow, sharp peak with full-width half-maximum (FWHM) of only
43\,$\mu$s, less than 1.5\% of the 2.95\,ms pulse period
(Fig.~\ref{fig:1909_profile}).  The brightness of the pulsar varies by
at least a factor of 30 in the 21-cm band. The mean flux density of
our observations is 3\,mJy, but this is biased by the fact that we tend
to observe longer when the pulsar is bright. Extremely high precision
timing observations are possible during episodes of favorable
scintillation.

\section{PULSE TIMING}
\label{sec:timing}

Following the confirmation of \psr, we began timing observations with
the $2 \times 512 \times 0.5$\,MHz Parkes filterbank centered on
1390\,MHz.  As a result of our incorrect discovery position, few
observations taken before February 2002 yielded data useful for
high-precision timing.  Integration times ranged from three to 100
minutes, with the longer observations broken into 10-minute
sub-integrations.  

We followed standard pulsar timing procedures: folded pulse profiles
from individual observations were cross-correlated with a high
signal-to-noise template profile to determine an average pulse time of
arrival (TOA), corrected to UTC(NIST).  The standard pulsar timing
package {\sc tempo}\footnote{http://pulsar.princeton.edu/tempo}, along
with the Jet Propulsion Laboratory's DE405 ephemeris, was used for all
timing analysis.  All 1.4\,GHz TOAs with uncertainty less than
1\,$\mu$s were included.  TOA uncertainties were multiplied by 1.25 to
achieve reduced $\chi^2 \simeq 1$.  Several observations were also
obtained in the 50-cm band with the $2 \times 256 \times 0.125$\,MHz
Parkes filterbank centered on 660\,MHz.  As these observations did not
yield TOA uncertainties less than 1\,$\mu$s they were used only to
determine the dispersion measure ($DM$), which was then held fixed.

\begin{figure}
\plotone{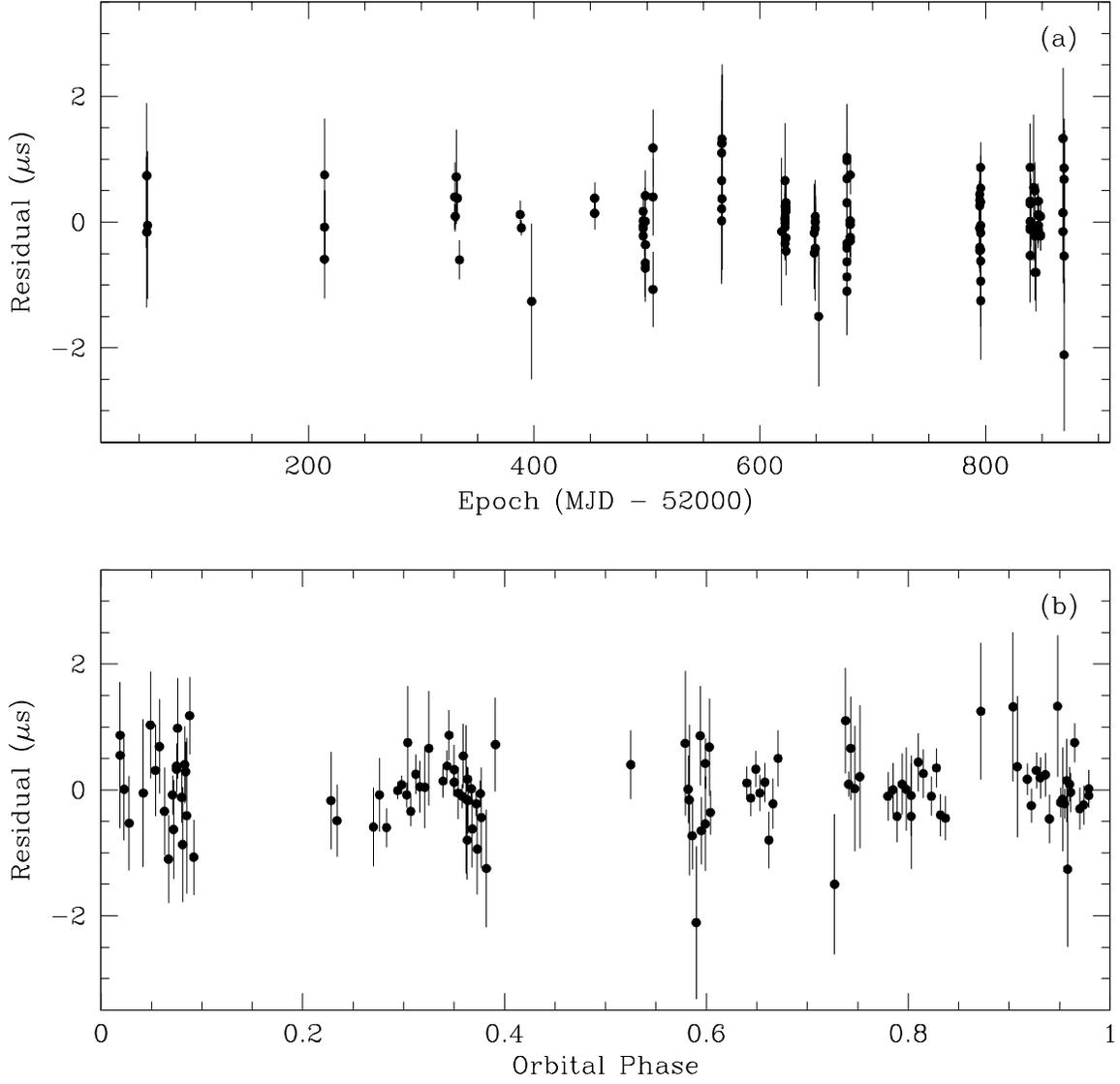}
\caption{Timing residuals for \psr.  (a): Residuals vs. observation epoch.  (b): Residuals vs. orbital phase.  Inferior conjunction occurs at orbital phase 0.16.}
\label{fig:1909_residuals}
\end{figure}

Because of the system's low eccentricity ($e$), we used the ELL1
binary model which replaces the familiar longitude of periastron
($\omega$), time of periastron ($T_0$), and $e$ with the time of
ascending node ($T_{\rm asc}$) and the Laplace-Lagrange parameters $e
\sin \omega$ and $e \cos \omega$ \citep{lcw+01}.  We give the results
of our timing analysis in Table~\ref{tab:par} and show residuals
relative to this best-fit model in Figure~\ref{fig:1909_residuals}.
The weighted RMS residual is only 330\,ns, among the best
obtained for any pulsar over $\sim$2 year timespans.

We have begun regular observations of \psr\ with the
Caltech-Parkes-Swinburne Recorder II \citep[CPSR2;][]{bai03}.  CPSR2
is a baseband recorder and real-time coherent dedispersion system with
two dual-polarization bands, each 64\,MHz wide.  The CPSR2 pulse
profile shown in Figure~\ref{fig:1909_profile} is smeared by only
2\,$\mu$s, compared with the 82\,$\mu$s time resolution of the
512-channel filterbank.  Initial results from CPSR2 data show great
promise for high precision timing, and will be reported in a future
paper.

\subsection{Shapiro Delay}

Though our current data set suffers from poor orbital phase coverage
(specifically a lack of high-quality observations near inferior
conjunction), we find strong evidence of Shapiro delay.  If we remove
Shapiro delay (i.e. companion mass, $m_{\rm c}$, and sine of orbital
inclination, $\sin i$) from the timing model and allow {\sc tempo} to
optimize the remaining parameters, $\chi ^2$ increases from 98.8 to
128.1 (for 102 and 104 degrees of freedom respectively), indicating a
highly significant detection.  The peak-to-peak residual is
essentially the same in either case (3.4\,$\mu$s) as much of the
Shapiro delay signal is absorbed into an erroneously large Roemer
delay; however, the weighted RMS residual increases to 370\,ns when
the model excludes $m_{\rm c}$ and $\sin i$.  Holding the astrometric,
orbital, and spin parameters fixed at the values in
Table~\ref{tab:par} but neglecting Shapiro delay increases the
peak-to-peak residual to 8.1\,$\mu$s, with pulses observed nearest inferior
conjunction arriving later than predicted by the model.

Given its large uncertainty, we are not concerned by the most likely
value of $\sin i > 1$ found by {\sc tempo}'s linear least-squares fit.
To confirm that the Shapiro delay is physically meaningful, we
constructed a $\chi ^2$ map as a function of $m_{\rm c}$ and $\sin i$,
while optimizing the remaining parameters.  Our sparse data set is
consistent with a large range of allowed parameter space, constraining
the system to have $\sin i \ge 0.85$ and $0.17\,{\rm M}_{\odot} \le
m_{\rm c} \le 0.55\,{\rm M}_{\odot}$ (95\% confidence).  The
covariance of the poorly-determined $m_{\rm c}$ and $\sin i$ with the
projected semimajor axis ($a \sin i$) and $e$ means that the
uncertainties in these parameters given in Table~\ref{tab:par} are
likely to be underestimated.

Currently, the strongest constraint on the companion mass is the
minimum mass of 0.195\,M$_{\odot}$ obtained from the mass function by
assuming $i = 90^\circ$ and $m_{\rm p} = 1.4$\,M$_{\odot}$; the fact
that Shapiro delay is detected at all suggests that we are likely
viewing the system roughly edge-on.  Future pulsar timing observations
near inferior conjunction will significantly improve the Shapiro
delay-derived companion mass.

\subsection{Distance}

We have measured the parallax, $\pi = 1.22\pm0.44\,$mas, and hence can
estimate the distance to \psr, $d_{\pi} = 0.82^{+0.46}_{-0.22}$\,kpc.
Combined with the proper motion, this distance corresponds to a
transverse velocity of $140^{+80}_{-40}$\,km\,s$^{-1}$.  The
$DM$-derived distance estimate is 0.46\,kpc \citep{cl02}.

The pulsar's proper motion induces an apparent radial acceleration
equal to $\mu^2 d / c$ \citep{shk70}.  This secular acceleration will
allow an entirely independent distance estimate from precise
measurement of the orbital period derivative, $\dot{P}_{\rm b}$
\citep{bb96}.  For circular binaries such as \psr, the intrinsic
$\dot{P}_{\rm b}$ due to general relativity is negligible.  However,
the secular acceleration will eventually produce a measurable
$\dot{P}_{\rm b}$ which can be used to determine the distance.  The
peak-to-peak timing signal due to parallax is $\sim\!1.4\,\mu$s, while
the $\dot{P}_{\rm b}$ signal increases with the observing baseline, $t$,
as $\sim\!240\,{\rm ns}\,(t / 1\,{\rm yr})^2$.  Thus, after only a few
years of timing $\dot{P}_{\rm b}$ will give the more precise distance.
We note that the acceleration induced by differential Galactic
rotation is about two orders of magnitude smaller than the secular
acceleration.

The secular acceleration also corrupts the observed period derivative
($\dot{P}$).  Since the pulsar's intrinsic spindown rate
($\dot{P}_{\rm int}$) is non-negative, we can obtain an upper distance
limit of 1.4\,kpc, consistent with the measured parallax.  Conversely,
the measured proper motion and parallax-derived distance allow us to
correct $\dot{P}$ for the secular acceleration and hence determine
$\dot{P}_{\rm int}$.

\section{OPTICAL OBSERVATIONS OF THE WHITE DWARF COMPANION}\label{sec:optical}

At the position of the pulsar, the digitized sky survey plates showed
a possible counterpart.  To verify this, we obtained images on the
night of 2003 June 4 with the Magellan Instant Camera (MagIC) at the
6.5-m Clay telescope on Las Campanas.  We took a 10-min exposure in $B$
and a 5-min exposure in $R$.  The conditions were poor, with thin clouds
and $1\arcsec$ seeing.

The images were reduced using the Munich image data analysis system
(MIDAS), following the usual steps of bias subtraction (separately for
the four amplifiers) and flat-fielding using dome flats.  For the
astrometric calibration, we selected 58 USNO-B1.0 \citep{mlc+03}
objects that overlapped with our images, were not overexposed, and
appeared stellar.  For these objects, we measured centroids and
fitted for zero-point position, plate scale, and position angle.
Rejecting 23 outliers with residuals larger than $0\farcs5$, the
inferred single-star measurement error in both bands is $0\farcs15$ in
each coordinate, which is consistent with expectations for the
USNO-B1.0 measurements.  Hence our astrometry should be tied to the
USNO-B1.0 system at the $\sim\!0\farcs03$ level.

\begin{figure}
\plotone{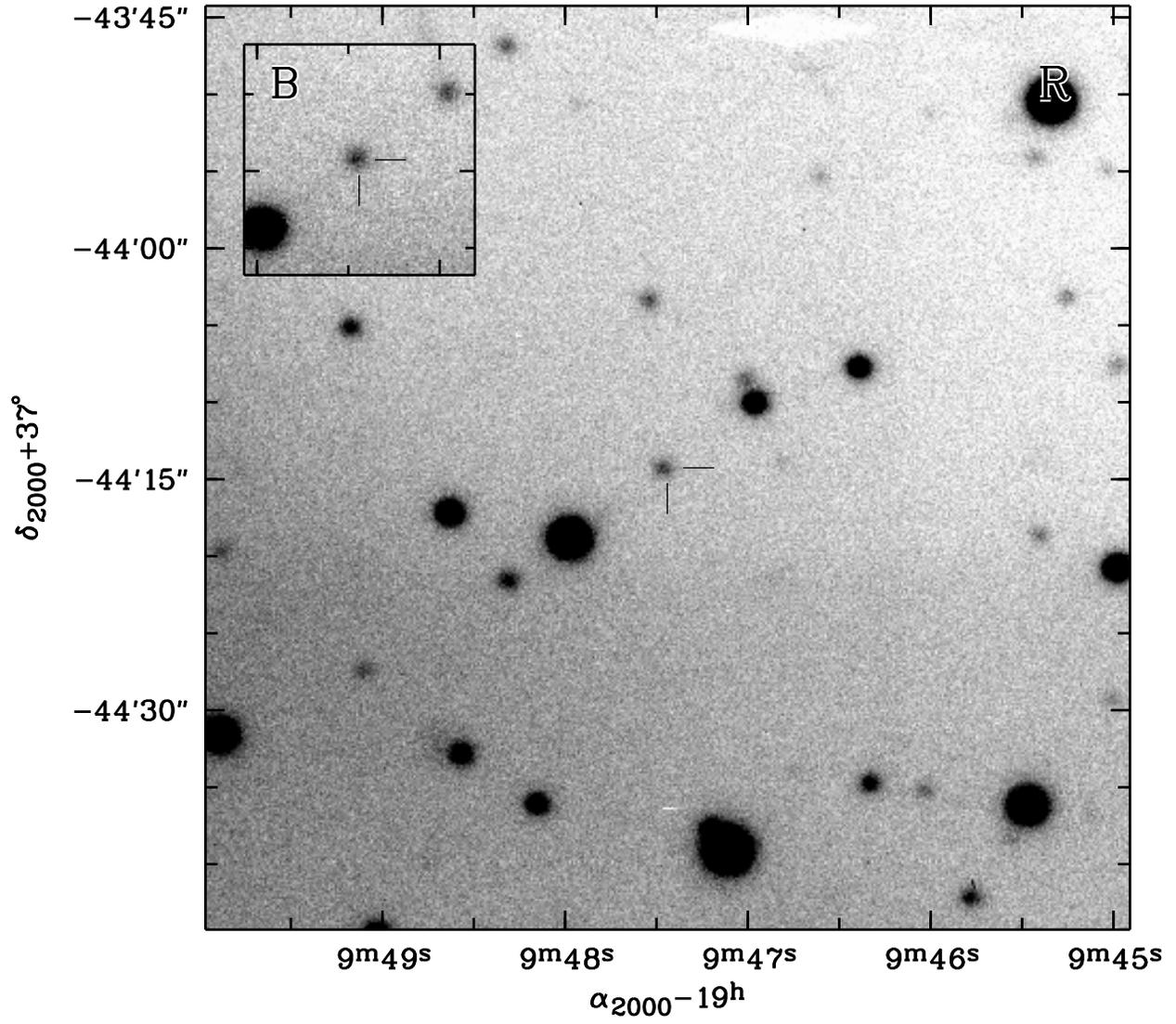}
\caption[]{Image of the field of \psr.  The large frame
shows our R-band image; the position of the pulsar is indicated by
the tick marks.  The inset shows the B-band image; compared to other
stars in the field, the companion is blue.\label{fig:image}}
\end{figure}

In both images, we found an object near the position of the pulsar
which is blue relative to other stars in the field; see
Fig.~\ref{fig:image}.  The position inferred from our astrometry is
$\alpha_{\rm J2000}=19^{\rm h}09^{\rm m}47\fs457$, $\delta_{\rm
J2000}=-37\arcdeg44\arcmin14\farcs17$, which is consistent with the
timing position listed in Table~\ref{tab:par} within the
uncertainty with which the USNO-B1.0 system is tied to the
International Celestial Reference Frame ($\sim\!0\farcs2$ in each
coordinate).

The identification of this object with the pulsar's companion was
confirmed by spectra, taken on the night of 2003 June 6, using the Low
Dispersion Survey Spectrograph~2 (LDSS2) at the Clay telescope.
We took two 30-min exposures with the high resolution grism and a
$1\arcsec$ slit, which gives a resolution of $\sim\!6\,$\AA.  The
conditions were mediocre, with thin cirrus and $0\farcs7$ seeing.  

\begin{figure}
\plotone{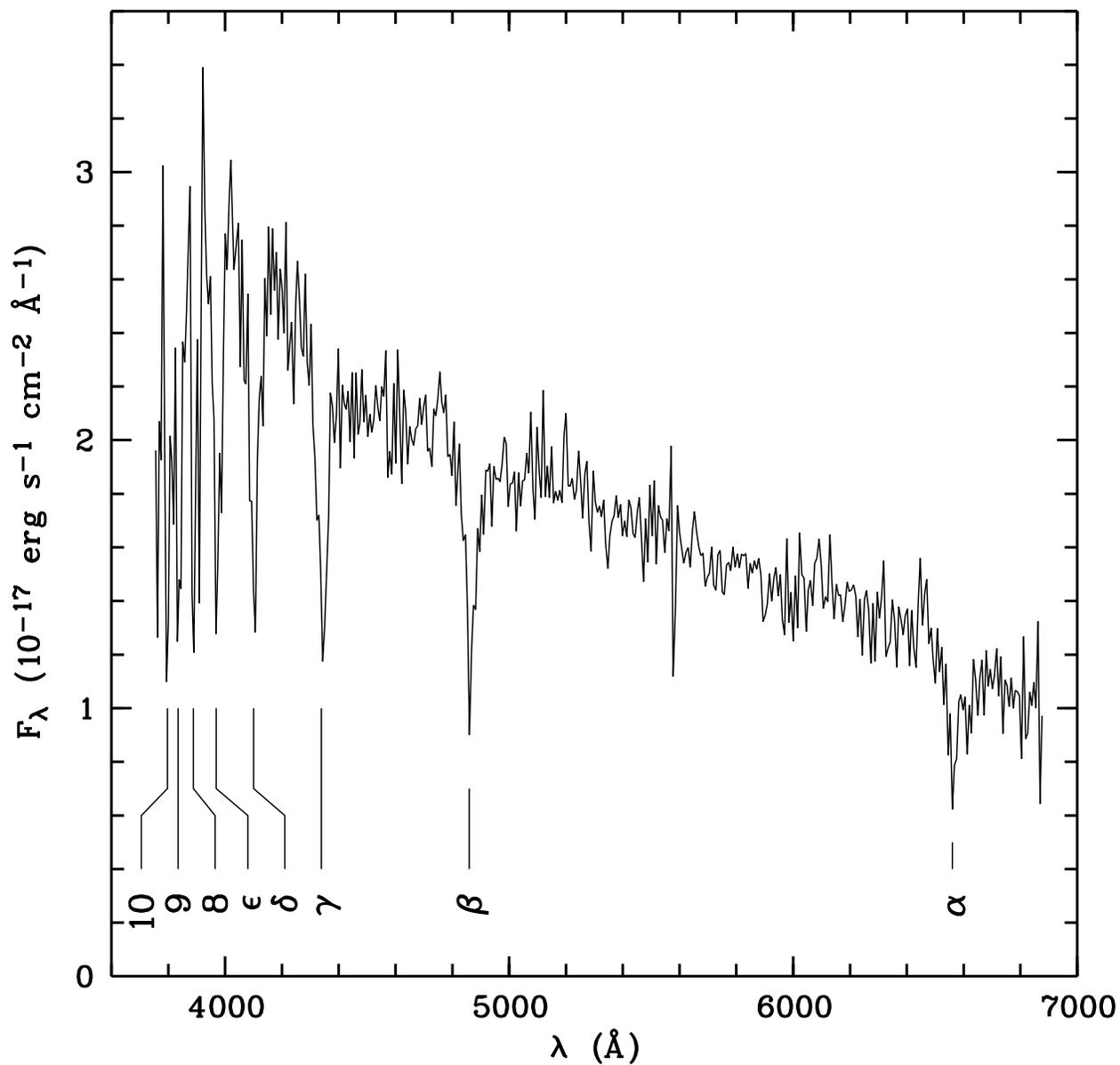}
\caption[]{Spectrum of the companion of PSR J1909$-$3744, showing that
it is a low-mass DA white dwarf.  The Balmer lines from H$\alpha$ to
H10 are indicated.  Note that the absolute flux calibration is
uncertain by about 50\%, as cirrus was present during the
observations.  The relative calibration, however, should be reliable
(except longward of $\sim\!6400\,$\AA, where the spectrum might be
contaminated by second-order light).\label{fig:spectrum}}
\end{figure}

We used MIDAS to reduce the spectra.  They were bias-subtracted
and flat-fielded using normalized dome flats (in which the noisy blue
part was set to unity).  After sky subtraction, the two spectra were
extracted using optimal weighting, and added together.  To derive the
response, we used spectra of Feige~110, taken during the same night,
and fluxes from calibrated STIS spectra.\footnote{See
http://www.stsci.edu/instruments/observatory/cdbs/calspec.html.}  The
result is shown in Fig.~\ref{fig:spectrum}.  We stress that since
cirrus was present, the absolute flux calibration is not reliable,
although both from the spectrum and from a rough photometric
calibration using USNO-B1.0 magnitudes, it follows that the companion
has $V\simeq21$.  

The spectrum (Fig.~\ref{fig:spectrum}) shows only the Balmer lines of
Hydrogen, from H$\alpha$ up to H10, consistent with a DA white dwarf
of low surface gravity (and thus low mass).  The strength of the lines
and the slope of the spectrum are similar to what is seen for PSR
J1012+5307 \citep{vkbk96} and PSR J0218+4232 \citep{bvkk03},
indicating a similar temperature of $\sim\!8500\,$K.\footnote{This
high temperature may be surprising, given the pulsar's long
characteristic age of roughly 8\,Gyr.  It indeed sets interesting
constraints on the thickness of the outer Hydrogen envelope of the
white dwarf; for a discussion and references, see \citet{bvkk03}.}  

Assuming a mass of $\sim\!0.20\,{\rm M}_\odot$, the white dwarf should
have a radius of approximately $0.03\,{\rm R}_\odot$ and a luminosity
of $\sim\!4\times10^{-3}\,{\rm L}_\odot$ \citep{dsbh98,rsab02}.
The corresponding absolute magnitude is $M_V\simeq 11.0$, which implies
$V\simeq 20.5$ at a distance of 0.8\,kpc, consistent with the rough
photometry quoted above.

\section{CONCLUSIONS}
\label{sec:conclusions}

Pulse timing precision is often limited by systematic errors that
distort the shape of pulsar profiles due to the interstellar medium
and imperfect polarimetric and instrumental response.  Intrinsically
narrow pulses are less susceptible to these errors, as are pulsars
with small dispersion measures.  \psr\ is likely to become a
cornerstone of future timing array experiments to detect low-frequency
gravitational waves. It is noteworthy that \psr\ is far from other
high-precision pulsars on the sky and therefore probes a unique line
of sight.

In the shorter run, further timing of the pulsar, especially with
CPSR2, should yield an accurate inclination and a
reasonably precise companion mass.  Furthermore, in a modest
spectroscopic observing campaign on an 8-m class telescope, the
radial velocity orbit of the companion can be measured precisely.
These combined results should allow a very reliable pulsar mass
measurement, and advance our understanding of binary pulsar evolution
and recycling.

\acknowledgments

We thank R. Edwards for invaluable help with pulsar search software,
H. Knight for assistance with Parkes observations, and M. Hamuy,
V. Mari\~no, and M. Roth for help with the Magellan observations.  BAJ
and SRK thank NSF and NASA for supporting their research.  MHvK
acknowledges support by the National Sciences and Engineering Research
Council of Canada.  The Parkes telescope is part of the Australia
Telescope which is funded by the Commonwealth of Australia for
operation as a National Facility managed by CSIRO.

\clearpage

\begin{deluxetable}{ll}
\tabletypesize{\scriptsize}
\tablewidth{0pt}
\tablecaption{Parameters of the \psr\ system.\label{tab:par}}
\tablehead{
\colhead{Parameter} & \colhead{Value\tablenotemark{a}}
}
\startdata
Right ascension, $\alpha_{\rm J2000}$\dotfill &  $19^{\rm h}09^{\rm m}47\fs44008(2)$  \\
Declination, $\delta_{\rm J2000}$\dotfill  &   $-37\arcdeg44\arcmin14\farcs226(1)$ \\
Proper motion in $\alpha$, $\mu_{\alpha}$ (mas yr$^{-1}$)\dotfill   &            $-$9.6(2) \\
Proper motion in $\delta$, $\mu_{\delta}$ (mas yr$^{-1}$)\dotfill  &           $-$35.6(7) \\
Annual parallax, $\pi$ (mas)\dotfill  &               1.22(44) \\
Pulse period, $P$ (ms)\dotfill & 2.9471080205034(1) \\
Period derivative, $\dot{P}$ (10$^{-20}$)\dotfill & 1.4026(3) \\
Reference epoch (MJD)\dotfill   &    52055.8704  \\
Dispersion measure, $DM$ (pc cm$^{-3}$)\dotfill       &       10.394(1) \\
Binary period, $P_{\rm b}$ (d)\dotfill     &    1.5334494503(1) \\
Projected semimajor axis, $a \sin i$ (lt-s)\dotfill      &      1.897992(4) \\
$e \sin \omega$ ($\times 10^{-7}$)\dotfill   &      0.1(90) \\
$e \cos \omega$ ($\times 10^{-7}$)\dotfill    &     $-$2.6(17) \\
Time of ascending node, $T_{\rm asc}$ (MJD)\dotfill  &   52055.87046667(6) \\
$\sin i$\dotfill    &          1.1(4) \\
Companion mass $m_{\rm c}$ (M$_{\odot}$)\dotfill    &          0.13(37) \\
\cutinhead{Derived Parameters}
Mass function, $f(m)$\dotfill &  0.00312196(2) \\
Orbital eccentricity, $e$ ($\times 10^{-7}$)\dotfill     &    2.6(18) \\
Longitude of periastron, $\omega$ (deg)\dotfill & $177.874428 \pm 190$ \\
Time of periastron, $T_0$\dotfill    &   $52056.6281373 \pm 0.8$\\
Parallax distance, $d_{\pi}$ (kpc)\dotfill & $0.82^{+0.46}_{-0.22}$ \\
Transverse velocity, $v_{\perp}$ (km s$^{-1}$)\dotfill & $140^{+80}_{-40}$  \\
Intrinsic period derivative, $\dot{P}_{\rm int}$ (10$^{-20}$)\tablenotemark{b}\dotfill & $0.61^{+0.23}_{-0.50}$ \\
Surface magnetic field, $B_{\rm surf}$ $(\times 10^8 \rm{G})$\tablenotemark{b}\dotfill & $1.3^{+0.2}_{-0.8}$ \\
Characteristic age, $\tau_{\rm c}$ (Gyr)\tablenotemark{b}\dotfill & $7.7^{+35.0}_{-2.1}$ \\
Galactic longitude, $l$ (deg)\dotfill & 359.73 \\
Galactic latitude, $b$ (deg)\dotfill &  $-$19.60 \\
Distance from Galactic plane, $|z|$ (kpc)\dotfill & $0.28^{+0.16}_{-0.07}$ \\
Pulse FWHM, $w_{50}$ ($\mu$s)\dotfill & 43 \\
Pulse width at 10\% peak, $w_{10}$ ($\mu$s)\dotfill & 89 \\
\enddata
\tablenotetext{a}{Figures in parenthesis are uncertainties in the last digit quoted.  Uncertainties are calculated from twice the formal error produced by {\sc tempo}.}
\tablenotetext{b}{Corrected for secular acceleration based on measured proper motion and parallax.}
\end{deluxetable}

\end{document}